\documentclass[aps,pra,a4paper, amsfonts, amssymb, amsmath, reprint, showkeys, nofootinbib, twoside]{revtex4-1}

\usepackage[english]{babel}
\usepackage[utf8]{inputenc}
\usepackage[colorinlistoftodos, color=green!40, prependcaption]{todonotes}
\usepackage{graphicx}
\usepackage{bm}
\usepackage{xcolor}

\begin{document}

\title{QED corrections to  Abraham and  Aharonov-Casher force on Rydberg atoms  }

\author{R. Le Fournis}
\email[]{romuald.lefournis@lpmmc.cnrs.fr}

\author{B.A. van Tiggelen}
\email[]{Bart.Van-Tiggelen@lpmmc.cnrs.fr}

\affiliation{Univ. Grenoble Alpes, CNRS, LPMMC, 38000 Grenoble, France}

\author{M. Donaire}
\email[]{manuel.donaire@uva.es}

\affiliation{Universidad de Valladolid, 47011 Valladolid, Spain}

\date{\today} 

\begin{abstract}
We calculate the Abraham force and the Aharonov-Casher force exerted by the electromagnetic quantum vacuum on alkaline atoms in highly excited, long-living Rydberg states.
Because of their high polarizability and long-life time such atoms are good candidates to observe these forces.
\end{abstract}


\maketitle

\section{Brief Context} \label{section1}

Electromagnetic fields exert forces on charge-neutral materials and are due to induced polarizations and magnetization inside the matter. From a practical point of view we can distinguish dissipative forces, dispersive forces and a class that we can summarize by Abraham forces. The first are due to absorption or scattering of electromagnetic waves and are very important at optical  frequencies.  Dispersive optical forces are governed by spatial gradients of electromagnetic energy. They are widely employed in optical tweezers and for the trapping of cold atoms. Abraham forces are created by time-dependence of the fields and can only be observed at low electromagnetic frequencies. They are small and controversial, but touch a fundamental issue in the interaction between matter and electromagnetic waves.

For over many decades, a so-called  ``Abraham-Minkowski" controversy has existed about the description of momentum of light in matter \cite{brevik_experiments_1979}.
This controversy stems from to the apparent ambiguity to identify the force exerted by light on matter, when relying on the macroscopic momentum conservation laws that govern the interaction between matter and electromagnetic field.

Several versions for the electromagnetic momentum of light in matter have emerged leading to different forces acting on material media.
The two most well-known are the Minkowski version, for which the momentum density $\bm{g}(\bm{r})$  is given by  $\bm{g}_M = \bm{D} \times \bm{B}/c_0$, and the Abraham version, for which  $\bm{g}_A = \bm{E} \times \bm{H}/c_0$, i.e. chosen such that it equals to the Poynting vector for energy flow divided by $c_0^2$.
The difference between the two  momenta can be written as
\begin{eqnarray}\label{AB}
    \Delta \bm{G} \equiv \bm{G}_M-\bm{G}_A &=&   \frac{1}{c_0}\int d^3\mathbf{r} \left(  \bm{P}\times\bm{B} +\bm{E}\times\bm{M}\right) \nonumber \\
    &=&  \frac{1}{c_0} \left(  \bm{d}\times\bm{B} +\bm{E}\times\bm{m}\right)
\end{eqnarray}

in terms of polarization density $\bm{P}$ and magnetization density $\bm{M}$. The second line, featuring total electric dipole and magnetic moment, applies when the applied fields vary little over the volume of the  matter, typically true at low frequencies. The force $d \Delta \mathbf{G} /dt $ highlights the Abraham-Minkowski controversy but is it really an observable force? Both terms in this equation are conveniently ``dual" in electric and magnetic response and we will refer to their time-derivatives as the Abraham force and the Aharonov-Casher force, respectively. Other versions have been proposed on the basis of different arguments \cite{brevik_experiments_1979}. Among these we mention the  Einstein-Laub version, insisting on a dual symmetry between electric and magnetic fields in the optical force, and the Nelson version that favors the momentum density $\bm{g}_N = \bm{E} \times \bm{B}/c_0 $ that emerges microscopically \cite{cohen-tannoudji_photons_2004}. The controversy is fostered  by the weakness of these forces and the experimental  difficulty to discriminate between different versions. 

Walker and Walker \cite{walker_measurement_1975} were the first to observe (part of) the Abraham term $( \partial_t\bm{d} )\times \bm{B}$ in a non-magnetic medium ($\mathbf{M}$ = 0), using an electric field oscillating at low frequencies, and a static orthogonal magnetic field. Their observation favored the Abraham version but was unable to distinguish between the Abraham version and the Einstein-Laub version, which would require magnetic response. A more recent experiment\cite{rikken_measurement_2011} observed the Abraham force with increased  accuracy and was designed to search for or to exclude the existence of even smaller QED corrections.

 On the theoretical side, many efforts have been undertaken to ``solve" the Abraham-Minkowski" controversy. The dual symmetry expressed by Eq.~(\ref{AB}) was emphasized by Hinds and
 Barnett \cite{hinds_momentum_2009} and deeper exploited. Bliokh and al. proposed a dual version of the electromagnetism Lagrangian to solve the problem \cite{bliokh_dual_2013}.  Dual symmetry is a remarkable property of Maxwell's equations in free space, but is broken as soon as electric and magnetic moments of matter are included. More specifically, the standard Maxwell equations involving non-relativistic matter do not predict an  Aharonov-Casher force $d/dt ( \mathbf{E} \times \mathbf{m} ) /c_0 $.  Only the force $-\partial_t\bm{E} \times \bm{m} /2c_0$ can be identified for a   molecule with a permanent magnetic moment $\mathbf{m}$ subject to an electric field slowly varying in time \cite{van_tiggelen_generalized_2019}. In a  gauge theory the Abraham momentum $\mathbf{d} \times \mathbf{B}/c_0$ of a charge-neutral molecule  emerges as the difference between kinetic and conserved pseudo-momentum. This puts  Eq.~(\ref{AB}) in a new perspective and assigns different roles for both the Abraham and Minkowski versions \cite{hinds_momentum_2009}. Gauge theory also associates Abraham momentum with the ''longitudinal" momentum  $- \sum_i q_i \mathbf{A}_i(\mathbf{r})
= \mathbf{d} \times \mathbf{B}/2c_0$ of the electromagnetic field in the Coulomb gauge. This establishes an interesting link with  the topological Aharonov-Bohm phase, even though the geometry to observe  the topological phase is designed to have no net force on the charge.  Gauge-invariance seems consistent with dual symmetry since the dual equivalent the Aharonov-Bohm phase, the so-called the  Aharonov-Casher (AC) phase involving a permanent magnetic moment rather than a net charge, has been predicted from elementary arguments  \cite{aharonov_topological_1984} and observed \cite{sangster_aharonov-casher_1995}. The AC momentum emerges naturally from the Dirac Hamiltonian with spin coupled to the electromagnetic field \citep{aharonov_topological_1984,sangster_aharonov-casher_1995}.

A number of studies discussed  the possible existence of an optical force associated with the AC momentum  $(1/c_0)\bm{E} \times \bm{m}$. Van Dam and Ruijgrok \cite{van_dam_classical_1980} showed that  an AC type force $(1/c_0)(g-2)\bm{m} \times \partial_t\bm{E}$ emerges in a relativistic, classical  description of a charged particle with spin $\mathbf{S}$ and magnetization $\mathbf{m}=g \mu_B \mathbf{S}$ where $g$ is the spin gyromagnetic ratio.
Another important contribution was made by Babiker and Horsley \cite{horsley_powerzienauwoolley_2006}. Using the fact that the center of mass $\mathbf{R}(t)$ of a set of moving particles with total rest mass $M$ is imposed by their energies and not by their rest masses, it is possible to derive an equation of motion that takes the simplified form
 \begin{equation}\label{Bab}
    M \ddot{ \bm{R}} = \frac{d}{dt}\left(   \frac{1}{c_0} \bm{d} \times \bm{B} +    \frac{1}{c_0}  \bm{E} \times \bm{m}\right) + \cdots
 \end{equation}
In this equation dispersive forces have been ignored, who are known to be dual. Duality has been restored, but since in special relativity $M \ddot{ \bm{R}}$ no longer equals the real force, it is inappropriate to speak about dual symmetry of the Abraham force. For that reason, the second term in Eq.\eqref{Bab} is sometimes referred  to as a ``hidden" force \cite{mansuripur_resolution_2010}, although the notion of ``hidden dual symmetry" is arguably more appropriate.

The Aharonov-Casher force has never been observed. To our knowledge, even the Aharonov-Casher topological phase has never been observed for a particle whose magnetic moment is entirely determined by its orbital angular momentum.
In this work we shall address two quantum aspects related to both forces in Eq.~(\ref{AB}). In former work \cite{van_tiggelen_qed_2012} we have demonstrated that the interaction of a hydrogen atom with the quantum vacuum provides a small QED correction to the Abraham momentum $\mathbf{d }\times \mathbf{B}/c_0$. We extend this approach to highly excited Rydberg atoms and investigate  how the QED correction to the Abraham force scales with their principal quantum number. Highly excited Rydberg atoms are interesting because they have large polarizabilities, behave almost hydrogen-like, and have large lifetimes.
We use the same formalism for hydrogen-like atoms put in a  quantum state with orbital angular momentum, exposed to a time-independent electric field. We show that their coupling to the electromagnetic quantum vacuum generates an Aharonov-Casher type of force.

\section{QED  Abraham force on excited  Rydberg states}
\label{section3}

In this section we discuss the Abraham force on an alkali-metal atom with a single electron excited in a circular Rydberg state. This discussion extends previous work \cite{van_tiggelen_qed_2012} on the QED correction of the Abraham force for a hydrogen atom in its ground state, exposed to a slowly varying, homogeneous electric field crossed with respect to a static, homogeneous magnetic field. Circular Rydberg with large orbits have a large electric polarizability and are good candidates to observe the classical Abraham force $d/dt (\bm{d} \times \bm{B}_0)/c_0$, also in view of their large life time. For a magnetic field with amplitude $B_0 = 1 \, \mbox{mT}$ oscillating at  frequency $\omega =10^4 \, \mbox{Hz}$  and
an electric field with $E_0=10^2 \,  \mbox{V/m}$, the force acting on a Rydberg atom with $n=50$ is roughly $10^3$ times larger than the Abraham force measured on atoms \citep{rikken_measurement_2011}. It is therefore relevant  to find out if QED corrections increase similarly. Rydberg atoms are relatively easy to describe since they behave hydrogen-like. This can be deduced from their energy spectrum that is well described by quantum defect theory \cite{gallagher_rydberg_1988},
\begin{equation*}
    E_{n \ell} = \frac{E_0}{(n-\delta_\ell)^2}
\end{equation*}
where $E_0$ is the ground state energy of the hydrogen atom. The quantum defect  $\delta_\ell$  is non-zero when a small overlap of the wave function exists with the core that has, unlike in atomic hydrogen, a finite size. The quantum defect is very small for large angular momentum $\ell$ because the excited state hardly overlaps with the core. For circular Rydberg states the quantum state is well approximated by the wave function of atomic hydrogen \citep{gallagher_rydberg_1988}.

In order to calculate the Abraham force on the atom we consider both an external electric field, that we choose to be time-independent, and an external homogeneous,
magnetic field that may slowly vary in time. We will replace the ionic core by a point charge with mass $m_1$, charge $+e$, and position $\mathbf{r}_1$. The outer electron is labeled by $\mathbf{r}_2$ and has mass $m_2$. The non-relativistic Hamiltonian which describes the interaction of the Rydberg atom with both external classical fields and the quantum electromagnetic vacuum is,
\begin{equation}\label{HA}
\mathcal{H} = \mathcal{H}_R(\bm{B}_0) + \mathcal{H}_S + W(\bm{B}_0)
\end{equation}
\begin{multline*}
    \mathcal{H}_R(\bm{B}_0) = \frac{1}{2 \mu} \left( \bm{p} + \frac{\Delta m}{M} \frac{e}{2c_0} \bm{B}_0 \times \bm{r} \right)^2 \\
    + \frac{1}{2M}\left(\bm{P} - \frac{e}{c_0} \bm{B}_0 \times \bm{r} \right)^2 - \frac{e^2}{r}
\end{multline*}
\begin{equation}
\mathcal{H}_S = -eE_0z
\end{equation}
Here $\bm{p} = \mu (\bm{p}_1/m_1-\bm{p}_2/m_2)$ denotes the internal canonical momentum and $\bm{P}=\bm{p}_1 + \bm{p}_2$ is the external canonical momentum related to the center of mass motion;
$M$ is the mass of the atom, $\Delta m = m_1 -m_2$ is the mass difference between the ionic core and the outer electron, $\bm{r}=\bm{r}_1 - \bm{r}_2$ is the relative position between the ionic core and the outer electron. The interaction with the quantum vacuum is,
\begin{multline}\label{WQV}
    W(\bm{B}_0) = -\frac{e}{c_0 m_1} \left( \bm{p} + \frac{m_1}{M} \bm{P} - \frac{e}{2} \bm{B}_0 \times \bm{r} \right) \cdot \bm{A}(\bm{r_1}) \\ -\frac{e}{c_0 m_2} \left( \bm{p} - \frac{m_2}{M} \bm{P} + \frac{e}{2} \bm{B}_0 \times \bm{r} \right) \cdot \bm{A}(\bm{r_2})
\end{multline}
where $\bm{A}(\bm{r})$ is the gauge field of the electromagnetic quantum field in the Coulomb gauge,
\begin{equation*}
    \bm{A}(\bm{r})= \sum_{\bm{k}\lambda} \mathcal{A}_{\bm{k}}\bm{\epsilon}_{\bm{k}\lambda}\left[a^{\dagger}_{\bm{k}\lambda}  e^{-i\bm{k}\cdot \bm{r}} + a_{\bm{k}\lambda} e^{i\bm{k}\cdot \bm{r}}\right]
\end{equation*}
with $\mathcal{A}_{\bm{k}} = \sqrt{2 \pi \hbar c_0 / k V}$ for a quantization volume $V$.
Because of gauge invariance, the interaction $W$ depends explicitly on the magnetic field. Although $m_1 \gg m_2$ we do not wish to make any approximation at this stage because we need both masses to apply mass renormalization.
A part from the momentum $\bm{P}$ canonical to the center of mass $\mathbf{R}$ of the atom, two other linear momenta exist.  The momentum $\bm{K}$, sometimes called pseudo-momentum \cite{nelson_momentum_1991}, commutes with $ \mathcal{H} $ and is conserved if the magnetic field is time-independent.
The kinetic momentum $\bm{P}^K = M \dot{\mathbf{R}} $ is associated with the movement of the atom. They are related by \cite{van_tiggelen_generalized_2019},
\begin{eqnarray}\label{KB}
    \bm{K}(\bm{B}_0) &= &\bm{P}^K + \frac{e}{c_0} \bm{B}_0 \times \bm{r} + \frac{e}{c_0}\Delta\bm{A} + \sum_{\bm{k} \lambda} \hbar \bm{k} a^{\dagger}_{\bm{k}\lambda} a_{\bm{k}\lambda}
    \nonumber \\
    &=& \bm{P} + \frac{e}{2c_0} \bm{B}_0 \times \bm{r} + \sum_{\bm{k} \lambda} \hbar \bm{k} a^{\dagger}_{\bm{k}\lambda} a_{\bm{k}\lambda}
\end{eqnarray}
with $\Delta\bm{A} = \bm{A}(\bm{r}_1) - \bm{A}(\bm{r}_2)$. The pseudo-momentum $\bm{K}_0(\bm{B}_0)$ of the atom alone is given by the  first two terms of the first line.  Its continuous eigenvalues are denoted by $\bm{Q}_0$. The last two terms of the first line represent longitudinal and transverse momentum of the quantized electromagnetic field \citep{cohen-tannoudji_photons_2004}.
Because in the second line, $\bm{K}$ depends \emph{explicitly} on the magnetic
field it is conserved only if the magnetic field is time-independent. Since the magnetic field is time-dependent here, the force on the atom will achieve an extra term $(1/2c_0)\partial_t \bm{B}_0 \times \langle \bm{d} \rangle$\citep{van_tiggelen_generalized_2019}. Any QED correction to this extra force is not of our interest here since it corresponds to a QED correction to the static polarizability included in its measured value.

A very specific feature of hydrogen-like excited states compared to the ground state is the degeneracy of levels.
In the presence of an electric field,  the parabolic basis, with quantum numbers $(n_1,n_2,m)$, is more adapted than the spherical basis $(n,l,m)$ since it diagonalizes both the Stark potential in a given degenerated subspace and the hydrogen Hamiltonian without the electric field  \citep{landau_quantum_2007}. The circular Rydberg state $\rvert nC \rangle \equiv \rvert n,l= n-1,m= n-1 \rangle$  is special because it coincides with the parabolic state $\rvert n_1= 0,n_2= 0,m=n-1 \rangle$ and is as such also an eigenstate in the degenerated subspace of the Rydberg Hamiltonian $\mathcal{H}_R$. Because $n_1=n_2$ it has no permanent dipole moment and therefore does not suffer from the torque $\bm{d} \times \bm{E}_0 $ that would make the angular momentum rotate in time. This simplifies perturbation theory, because the electric field along the $z$-axis only couples states with the same quantum number $m$.  Since
 $n_1+n_2+|m|+1=n$ inside the subspace of $E_n$, no other states exist inside this subspace that also have $m =n-1$. In addition lower subspaces of $E_{n'}$ with $n' < n$ do not have this large value of $m$.

 We first ignore the quantum vacuum, so that the only perturbations are   $\mathcal{H}_S= -e E_0 z$ and the Zeeman interaction $\mathcal{H}_Z=- B_0 m_x$ with the magnetic moment  $m_x = -(e/2 m_e c_0)L_x$ for a magnetic field in the $x$ direction. These operators  do not commute which complicates a general analysis.  We shall make the realistic assumption that the Zeeman interaction is small compared to the Stark interaction. This imposes $B_0 \ll E_0 n / \alpha$, with $\alpha$ the fine structure constant. In this picture, the magnetic field changes the parabolic states without affecting the degenerate perturbation theory to cope with the electric field. In that case  we can first perturb the atomic state with the Stark interaction and next deal with the Zeeman term. Perturbation of the circular Rydberg state by the electric field gives,

\begin{multline}
\rvert nC,\bm{E}_0 \rangle  = \rvert nC \rangle + \sum_{j \notin \mu(E_n)}\rvert j \rangle \frac{\langle j \lvert \mathcal{H}_S \rvert nC \rangle}{E^{(0)}_{nC} - E^{(0)}_j}
      + \cdots
\label{electricexpansion2}
\end{multline}

Since $m$ is conserved by $\mathcal{H}_S$ , the sum excludes any other state inside the degenerated subspace and non-degenerate perturbation theory applies. The small parameter in this perturbation expansion is $E_S(n)/E_{\mathrm{Coulomb}}(n) \sim (E_0/ E_{at})n^5$ with $E_{at} = e/a_0^2 \approx 5\cdot 10^{11}$ V/m the atomic unit of the electric field. For $n=50$, and an electric field of $10^2$ V/m this ratio equals $0.06$.

With this perturbed state we can first recover the classical Abraham momentum in Eq.~(\ref{AB}) from the second term in Eq.~(\ref{KB}),
\begin{eqnarray}
    \langle \bm{P}_A \rangle  &=&  -\frac{e}{c_0} \langle nC, \bm{E}_0 \lvert  \bm{B}_0 \times \bm{r} \rvert nC, \bm{E}_0 \rangle \nonumber \\
    &=& \frac{1}{c_0} \alpha_{zz}(0)\bm{E}_0 \times \bm{B}_0
\label{formulestaticpolarizability}
\end{eqnarray}
with the static electronic polarizability of the circular Rydberg state \citep{landau_quantum_2007} given by,
\begin{eqnarray}
    \alpha_{zz}(nC) &=& 2 e^2 \sum_{j \notin \mu_n}\frac{\lvert \langle nC \lvert z \rvert j \rangle \lvert^2 }{E_j-E_{nC}} \nonumber \\
    &=& \frac{1}{8}n^4(17n^2-9m^2+19)a^3_0
    \label{Staticpolarizability}
\end{eqnarray}
here $m=n-1$. This expression is derived for atomic hydrogen, but applies to excited Rydberg states as well because the sum involves only states far from the ionic core with no quantum defects. We  also verified that for large  $n$, the sum over only all bound states in Eq.~(\ref{Staticpolarizability}) is very close to the exact result in the second line, meaning that ionized continuum states $ \rvert j \rangle $ hardly contribute to the static polarizability of the Rydberg state.

Before we calculate the expectation value of the momentum of the electromagnetic quantum vacuum in Eq.~(\ref{KB}),  we first need to perturb the Stark eigenfunction by the Zeeman interaction,
\begin{eqnarray}
 \rvert nC,\bm{E}_0, \bm{B}_0 \rangle  =\rvert nC, \bm{E}_0  \rangle
  & +&  \sum_{j \ne nC}\rvert j, \bm{E}_0 \rangle\frac{\langle j, \bm{E}_0 \lvert \mathcal{H}_Z \rvert nC, \bm{E}_0 \rangle}{E_j - E_{nC}}\nonumber\\
&+& \cdots \label{electricmagneticexpansion}
\end{eqnarray}
where $E_j$ is the energy of the state $\rvert j, \bm{E}_0 \rangle$  and $E_{nC}$ is the energy of the unperturbed state $ \rvert nC, \bm{E}_0 \rangle$. Part of the sum in this expansion involves states inside the originally degenerated subspace $\mu(E_n)$ that have a permanent dipole moment and thus achieve a linear Stark shift $E_S = \frac{3}{2}n(n_1 - n_2)E_0 a_0$  from the original level $E_{nC}$.  Some of them are coupled by the Zeeman interaction. The energies of the states $\rvert n_1, n_2, m, \bm{E}_0 \rangle$ and $\rvert n_1, n_2, -m, \bm{E}_0 \rangle$ in the subspace  $\mu(E_n)$ are still degenerated, but are not coupled by the Zeeman effect, so that no singularities exist in the sum over the states $\rvert j, \bm{E}_0 \rangle$ inside the subspace.  The modification of the eigenfunction $\rvert {nC}, \bm{E}_0\rangle$ by the Zeeman effect is small because the Zeeman interaction was assumed small compared to the Stark interaction.

We shall couple the magneto-electric circular Rydberg state $\lvert nC , \bm{Q}_0,\bm{E}_0,\bm{B}_0 \rangle\equiv \lvert \{a(nC)\} \rangle $ to the electromagnetic quantum vacuum in the usual way. Here $\bm{Q}_0$ is quantum number of the pseudo-momentum $\bm{K}_0 $ the Rydberg atom in Eq.~(\ref{KB}). We define $\lvert \{a\}  \lbrace n \rbrace \rangle \equiv \lvert  \{a\} \rangle  \otimes
\lvert \lbrace n \rbrace \rangle$  as the direct product states of the  atomic  states and the Fock states with photon occupation   $n_\mathbf{k}$
  and energy $E_{\lbrace n \rbrace } = \Sigma_\mathbf{k} n_\mathbf{k} \hbar \omega_\mathbf{k}$.
In the standard approach \cite{loudon_quantum_2000} one slowly switches on the interaction with the quantum vacuum as $W (t) = \exp(\eta t /\hbar)W $. At $t_0 \rightarrow - \infty$  the wave function of atom plus photon field is chosen to be a direct product of empty photon Fock states $\lbrace 0 \rbrace \rangle$  and the magneto-electric Rydberg state with kinetic momentum $\mathbf{Q}_0$.  According to time-dependent perturbation theory the wave function of the circular state behaves as,

\begin{multline}
    \lvert \Psi_{\{a(nC)\}}  \rangle =  \exp\left(-{i} \int_{t_0}^t \Delta_{nC} dt' \right) \lvert \{a\} \lbrace 0 \rbrace \rangle \\ +\sum'_{\{a'\} \lbrace n \rbrace} \frac{W(\bm{B}_0)_{ \{a'\} \lbrace n \rbrace,nC \lbrace 0 \rbrace}}{E_{\{a(nC)\} \lbrace 0 \rbrace} - E_{ \{a'\} \lbrace n \rbrace}+i\eta} \lvert \{a'\}  \lbrace n \rbrace \rangle
    \label{vacuumexpansion2}
\end{multline}
The accent ($'$) reminds us not to sum over the unperturbed state $\rvert \{a(nC)\}  \lbrace 0 \rbrace \rangle$.  The complex frequency  $\Delta_{nC}$ denotes the Lamb shift and the spontaneous decay rate to states with lower energy.
Recall that we search for a momentum linear in magnetic field. We can apply \emph{mutadis mutandis} previous work done for hydrogen that  identified after mass renormalization two different contributions to the expectation value of the longitudinal momentum $e \Delta\mathbf{ A}/c_0$ of the quantum vacuum featuring in the pseudo-momentum~(\ref{KB})\citep{van_tiggelen_qed_2012}. QED corrections to the Abraham-momentum $ {e/c_0} \bm{B}_0 \times \bm{r}$ exist as well but will be considered as corrections to the static polarizability and as such part of the classical Abraham momentum.
  In the first contribution - $\langle \bm{P}_{\mathrm{long}} \rangle_1$ -  the magnetic field enters via the Zeeman splitting of the atomic Hamiltonian, in the second - $\langle \bm{P}_{\mathrm{long}} \rangle_2$ - the magnetic field enters via the gauge operators $e\bm{B}_0 \times \bm{r}/2m_i$ of the interaction $W(\bm{B}_0)$ with the quantum vacuum, given by
  Eq.~(\ref{WQV}). To calculate the second, it is enough to  use the perturbation \eqref{electricexpansion2} in the electric field, whereas the first requires the  expansion \eqref{electricmagneticexpansion} in the magnetic field. Both contributions point either along or opposite to
$ \bm{E}_{0} \times \bm{B}_{0} \sim \bm{P}_A$ because the quantum state of the atom has its angular momentum chosen along the electric field so that there cannot be a third direction in this problem. We explicitly checked that for an atomic angular momentum chosen \emph{opposite} to the electric field, the same direction for $\langle \bm{P}_{\mathrm{long}} \rangle$ is obtained as required by time-reversal symmetry.
We obtain,
\begin{equation*}
    \langle \bm{P}_{\mathrm{long}} \rangle = \langle \bm{P}_{\mathrm{long}} \rangle_1 + \langle \bm{P}_{\mathrm{long}} \rangle_2
\end{equation*}
with
\begin{eqnarray}
\langle \bm{P}_{\mathrm{long}} \rangle_1 &= & \frac{1}{2}\alpha^2 a_0 \langle nC, \bm{E}_0,\bm{B}_0 \lvert \Delta(\hat{\bm{r}})\cdot \bm{p} \rvert  nC, \bm{E}_0,\bm{B}_0 \rangle \nonumber \\
&\equiv & \alpha^2 \kappa_1(n) \bm{P}_A
\end{eqnarray}
where $\Delta(\hat{\bm{r}})= r^{-1}(1+\hat{\bm{r}}\hat{\bm{r}})$ is an operator that emerges from the integration over the wave number $\bm{k}$ of the virtual photon.
The second contribution is
\begin{eqnarray}
\langle \bm{P}_{\mathrm{long}} \rangle_2 &=& \frac{e}{2}\alpha^2 a_0 \langle nC, \bm{E}_0 \lvert \bm{B}_0 \times \hat{\bm{r}} \rvert  nC, \bm{E}_0 \rangle
\label{Plong2} \nonumber \\
&\equiv & \alpha^2 \kappa_2(n) \bm{P}_A
\end{eqnarray}
We start with the calculation of the first contribution. Inserting the Stark wave function \eqref{electricmagneticexpansion} leads to
\begin{multline}
\langle \bm{P}_{\mathrm{long}} \rangle_1 = \frac{1}{2}\alpha^2 a_0 \sum_{j \ne nC}\langle nC, \bm{E}_0 \lvert \Delta(\hat{\bm{r}}) \cdot \bm{p} \rvert j,\bm{E}_0 \rangle
\\
\times \frac{1}{E_{nC}-E_j} \langle j,\bm{E}_0 \lvert \mathcal{H}_Z \rvert nC, \bm{E}_0 \rangle + c.c
\label{Plong1magneticexpansion}
\end{multline}

It is convenient to separate this sum into \emph{a)}  a sum inside the original degenerate subset  $j \in \mu(E_n)$ and \emph{b)} a sum over all states outside $j \notin \mu(E_n)$.  The first case is characteristic for the Rydberg atoms and is absent  for hydrogen in its ground state. For the case  \emph{b)} the energy level separation of quantum states in different subsets is assumed to be much larger than the Stark shift (which excludes too large values for $n$ in  which case other complications arise such as ionization). Since $[\mathcal{H}_R,\mathcal{H}_Z] =0$, the matrix element $\langle j \lvert \mathcal{H}_Z \rvert nC \rangle =0$ for any state $ \rvert j \rangle $
outside $ \mu(E_n)$. The matrix element $ \langle j,\bm{E}_0 \lvert \mathcal{H}_Z \rvert nC, \bm{E}_0 \rangle$ is thus at least linear in
$\bm{E}_0$ so that we can ignore the Stark shifts in $E_{nC}$ and $E_j$ in Eq.~\eqref{Plong1magneticexpansion},  as well as
the electric field dependence of the first matrix element. This gives,

 \begin{multline}
\langle \bm{P}_{\mathrm{long}} \rangle_{1b} = \frac{1}{2}\alpha^2 a_0 \sum_{j \notin \mu(E_n)}\langle nC \lvert \Delta(\hat{\bm{r}}) \cdot \bm{p} \rvert j \rangle
\\
\times \frac{1}{E^{(0)}_{nC}-E^{(0)}_j} \langle j,\bm{E}_0 \lvert \mathcal{H}_Z \rvert nC, \bm{E}_0 \rangle + c.c
\label{Plong1out}
\end{multline}

\begin{figure}
\begin{center}
\hspace*{-1cm}\includegraphics[scale=0.6]{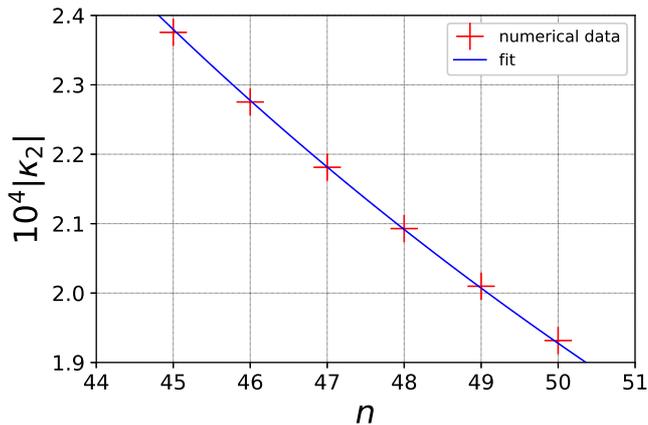}
\end{center}
\caption{Absolute value of the parameter $\kappa_2(n)$, describing the QED correction to the Abraham momentum via the magneto-optical interaction with the quantum vacuum, as a function of the principal quantum number $n$ of the Rydberg state. The fit shows that $\kappa_2(n)$ varies  as $n^{-2}$.}
\label{fig:kappa2fig}
\end{figure}
\begin{figure}
\begin{center}
\hspace*{-1cm}\includegraphics[scale=0.6]{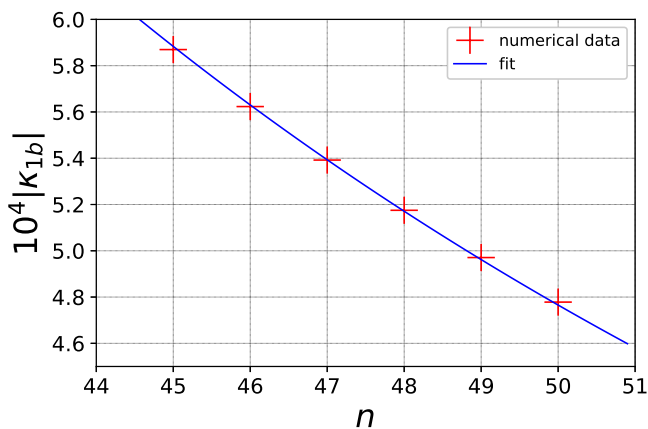}
\end{center}
\caption{Absolute value of the parameter $\kappa_{1b}(n)$, describing the QED correction to the Abraham momentum via the Zeeman splitting of the atomic energy levels outside the subspace associated with the level  $\rvert nC,\bm{E}_0 \rangle $, as a function of the principal quantum number $n$. The fit shows that $\kappa_{1b}(n)$  varies  as $n^{-2}$.}
\label{fig:kappa1outfig}
\end{figure}
\begin{figure}
\begin{center}
\hspace*{-1cm}\includegraphics[scale=0.6]{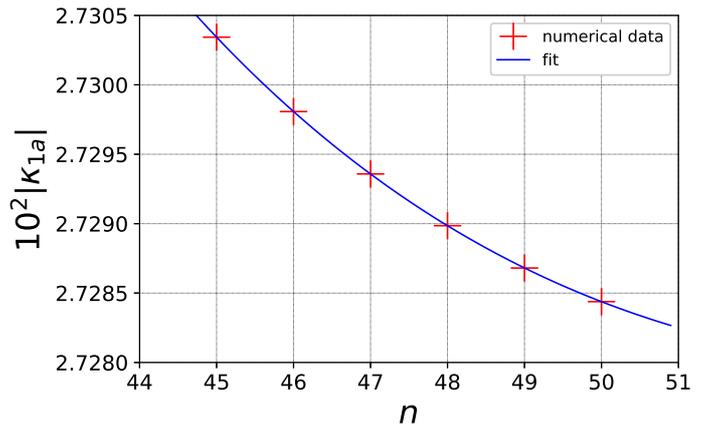}
\end{center}
\caption{Absolute value of the parameter $\kappa_{1a}(n)$, describing the QED correction to the Abraham momentum via the Zeeman splitting of the atomic energy levels inside the subspace associated with the level  $\rvert nC,\bm{E}_0 \rangle $, as a function of the principal quantum number $n$. The fit shown is $f(n)=2.78 + 5,788\times 10^{-2}\left(50/n\right)^2-1.05 \times 10^{-1}(50/n)$.}
\label{fig:kappa1infig}
\end{figure}

This only depends on the Stark eigenfunction $\rvert j,\bm{E}_0 \rangle$ whose correction linear in the electric field is given by Eq.~(\ref{electricexpansion2}).

For states inside the subspace $ \mu(E_n)$ the energy difference in the denominator of Eq.\eqref{Plong1magneticexpansion} is of the order of the Stark shift. We can collect terms proportional to   $1/E_0$ and $E_0$
by using higher-order perturbation theory for the Stark eigenfunctions and eigenvalues,
\begin{multline}
\langle \bm{P}_{\mathrm{long}} \rangle_{1a} = \\ \frac{1}{2}\alpha^2 a_0 \biggl[\sum_{j \ne nC}^{j \in \mu(E_n)}
\frac{\langle nC, \bm{E}_0 \lvert \Delta(\hat{\bm{r}}) \cdot \bm{p} \rvert j,\bm{E}_0 \rangle \langle j,\bm{E}_0 \lvert \mathcal{H}_Z \rvert nC, \bm{E}_0 \rangle}{E^{(1)}_{nC}-E^{(1)}_j}   \\ + (E^{(2)}_n-E^{(2)}_j) \frac{\langle nC, \bm{E}_0 \lvert \Delta(\hat{\bm{r}}) \cdot \bm{p} \rvert j,\bm{E}_0 \rangle}{(E^{(1)}_{j}-E^{(1)}_{nC})} \frac{\langle j,\bm{E}_0 \lvert \mathcal{H}_Z \rvert nC, \bm{E}_0 \rangle}{E^{(1)}_{nC}-E^{(1)}_j} \\ + (E^{(3)}_n-E^{(3)}_j) \frac{\langle nC \lvert \Delta(\hat{\bm{r}}) \cdot \bm{p} \rvert j \rangle}{(E^{(1)}_{j}-E^{(1)}_{nC})} \frac{\langle j \lvert \mathcal{H}_Z \rvert nC \rangle}{E^{(1)}_{nC}-E^{(1)}_j}\\
+ (E^{(2)}_n-E^{(2)}_j)^2 \frac{\langle nC \lvert \Delta(\hat{\bm{r}}) \cdot \bm{p} \rvert j \rangle}{(E^{(1)}_{j}-E^{(1)}_{nC})^2} \frac{\langle j \lvert \mathcal{H}_Z \rvert nC \rangle}{E^{(1)}_{nC}-E^{(1)}_j} + c.c \biggr]
\label{kappa1in}
\end{multline}

Except for the first term,  all terms in this expansion are of the same order but behave  differently as $n$ increases.
The first term generates a momentum proportional to $(1/E_0)\bm{B}_0 \times \hat{\bm{z}} $. Because the Rydberg state was given an angular momentum along the $\mathbf{z}$-axis the matrix element in the numerator of the first term survives even when $\bm{E}_0=0$.
By choosing the angular momentum opposite to $\bm{E}_0$ we verified  that the same direction is found, consistent with time-reversal symmetry.  It decays rapidly as  $n^{-4}$ whereas all others grow with $n$.   For a typical electric field $E_0 \sim 10^2 \, \mbox{V/m}$ and $n=50$ it is much smaller than the other terms. Among all terms in Eq.~(\ref{kappa1in}), the dominating contribution is  obtained by applying second-order perturbation of  the Stark wave function in the first term that we will not specify in detail. This compensates the factor $E_0$ in the denominator, and  produces a momentum of the form
$\bm{E}_0 \times \bm{B}_0$.

We have evaluated Eqs.~\eqref{Plong1out} and \eqref{kappa1in} numerically and  have omitted the negligable term proportional to $(1/E_0)\bm{B}_0 \times \hat{\bm{z}}$ in the first term of Eq.\eqref{kappa1in}. The results are shown in Figs.~[\ref{fig:kappa2fig}-\ref{fig:kappa1infig}]. We have included only the eigenfunctions associated with the discrete spectrum. For the ground state the continuum makes a relevant contribution \cite{van_tiggelen_qed_2012}, but like was mentioned earlier to be the case for the static polarizability~(\ref{Staticpolarizability}) we expect the contribution of the continuous spectrum to be negligible. The momenta  $\langle \bm{P}_{\mathrm{long}} \rangle_{1a}$ and $\langle \bm{P}_{\mathrm{long}} \rangle_{2}$ point in the same direction as $\bm{P}_A$ whereas the correction $\langle \bm{P}_{\mathrm{long}} \rangle_{1b}$ is in the opposite direction. Figure \ref{fig:kappa1infig} shows that the correction $\kappa_{1a}(n) \sim a + b \times (50/n) + c \times (50/n)^2$ dominates over the corrections $\kappa_2$ and $\kappa_{1b}$.
We conclude that the quantum vacuum correction to the classical Abraham momentum of a highly excited  Rydberg state is positive and of relative order $0.028 \alpha^2$.  The same order of magnitude was found for the ground state of hydrogen, that  has however a much smaller classical Abraham momentum.

\section{QED Aharonov-Casher type force on a Rydberg atom} \label{section2}
In this section we consider an atom in an single excited Rydberg state exposed to only a homogeneous electric field.
The electric field is assumed time-independent, so that no classical optical force of the kind $  \mathbf{m} \times (d\mathbf{E}_0/dt)/c_0$ is exerted on the atom \cite{van_tiggelen_generalized_2019}.
Contrary to the previous section, the atom is assumed to be in a quantum state associated with an angular momentum $\mathbf{L}$ that is not aligned with the electric field. This geometry  is not conserved in time  and $\mathbf{L}(t)$ will either oscillate or rotate in a plane perpendicular to the applied electric field \cite{landau_quantum_2007}. In the following we will derive an Aharonov-Casher force proportional to $  (d\mathbf{m}/dt)  \times \mathbf{E}/c_0$ caused by the interaction with the quantum vacuum.

If we couple this atom to the electromagnetic field,  the non-relativistic Hamiltonian $\mathcal{H}$ of atom and quantized radiation in Eq.~(\ref{HA}) simplifies to
 \begin{equation}\label{HAC}
    \mathcal{H} = \mathcal{H}_{\mathrm{at}} + W + \mathcal{H}_F
\end{equation}
with
\begin{equation*}
   \mathcal{H}_{\mathrm{at}}   = \frac{\bm{p}^2}{2 \mu} + \frac{\bm{P}^2}{2M} - \frac{e^2}{r}+  \mathcal{H}_S,
\end{equation*}
with the Stark interaction and the transverse photon Hamiltonian given by
\begin{equation*}
     \quad \mathcal{H}_S = -e E_0 z, \quad \mathcal{H}_F = \sum_{\bm{k}\lambda} \hbar \omega_{\bm{k}} a^{\dagger}_{\bm{k}\lambda} a_{\bm{k}\lambda},
\end{equation*}
and the interaction with the quantum vacuum,
\begin{equation*}
     W = \frac{e}{c_0 m_1}\left(-\bm{p}-\frac{m_1}{M}\bm{P}\right) \cdot \bm{A}(\bm{r}_1)  -\frac{e}{c_0 m_2}\left(\bm{p}-\frac{m_2}{M}\bm{P}\right) \cdot \bm{A}(\bm{r}_2)
\end{equation*}
Particle $1$ is the Rydberg ion with charge $+e$, particle $2$ is the electron.
 We choose the external electric field again along the $z$-axis. Photonic operators $\mathbf{A}^2$ have been ignored.  The pseudo-momentum defined in Eq.~(\ref{KB}) becomes
\begin{equation}\label{K}
    \bm{K} = \bm{P}^K + \frac{e}{c_0}\bm{A}(\mathbf{r}_1) -  \frac{e}{c_0}\bm{A}(\mathbf{r}_2) + \sum_{\bm{k} \lambda} \hbar \bm{k} a^{\dagger}_{\bm{k}\lambda} a_{\bm{k}\lambda}
\end{equation}
If $\rvert \Psi(t) \rangle$ is the quantum state of atom and electromagnetic field at time $t$, the expectation value
$\langle \bm{K} \rangle = \langle \Psi (t) \lvert \bm{K} \rvert \Psi(t) \rangle$
can be separated into a kinetic, an electromagnetic longitudinal and an electromagnetic transverse momentum. Since $\mathbf{K}$ is conserved in time we find for the force on the atom,
\begin{equation*}
   \bm{F} = \frac{d \langle \bm{P}^K \rangle}{dt} = -  \frac{d }{dt}\biggl[\langle \bm{P}_{\mathrm{long}} \rangle + \langle \bm{P}_{\mathrm{trans}} \rangle \biggr]
\end{equation*}
Without the quantum vacuum, this force vanishes since the atom is neutral and no classical Aharonov-Casher force exists. QED corrections can come from both terms in this expression, but we find the transverse part typically a factor $\alpha^2 = (e^2/\hbar c_0)^2$ smaller than the longitudinal momentum (see Appendix A). Because we cannot concentrate on a single circular Rydberg state with angular momentum directed along the electric field (in which case the Aharonov-Casher momentum $\bm{m} \times \bm{E}_0$ would vanish) we have to be more careful with degenerated perturbation theory of the Stark interaction  than in the previous section.
A parabolic state $\rvert \ell \rangle = \rvert n_1,n_2,m \rangle$ in the subspace $\mu_\ell $,  defined by the set of eigenfunctions of $\mathcal{H}_{\mathrm{at}}$ with equal eigenenergy is linearly perturbed by the electric field according to \citep{landau_quantum_2007}
\begin{multline}
    \lvert \ell \bm{E}_0 \rangle =    \rvert \ell \rangle +  \sum_{n \notin \mu_\ell} \frac{\langle n \lvert \mathcal{H}_S \rvert \ell \rangle }{E^{(0)}_\ell - E^{(0)}_n} \rvert n \rangle \\
     + \sum^{E^{(1)}_{\ell} \ne E^{(1)}_{\ell'}}_{\ell' \in  \mu_\ell \ne \ell }\sum_{n \notin \mu_\ell} \frac{\rvert \ell' \rangle}{E^{(1)}_{\ell'} - E^{(1)}_{\ell}}\frac{\langle \ell' \lvert \mathcal{H}_S \rvert n \rangle \langle n \lvert \mathcal{H}_S \rvert \ell \rangle}{E^{(0)}_\ell - E^{(0)}_n}
     \label{electricfieldexpansion}
\end{multline}
with  $E^{(1)}_{\ell} = \langle \ell \lvert \mathcal{H}_S \rvert \ell \rangle $ the first-order Stark shift of the parabolic state $\rvert \ell \rangle$. If $\rvert \ell \rangle $ is not a circular state, the third complicated term does not necessarily vanish.

We next add the quantum vacuum using the same procedure as in the previous section \cite{loudon_quantum_2000}. This time,   the wave function of atom plus photon field at $t_0 \rightarrow - \infty$  is chosen to be a direct product of empty photon Fock states $\lbrace \ell \bm{E}_0 \rbrace \rangle$  and some superposition of parabolic states $\sum_\ell \beta_\ell \rvert  \ell  \bm{E}_0 \rangle $ with same energy $E_\ell^{(0)}$ and same kinetic momentum $\mathbf{Q}_0$. The coefficients $\beta_\ell$ will later be chosen such that this state has a well-defined orbital momentum and remains Rydberg-like in time. According to time-dependent perturbation theory, the wave function behaves as,
\begin{multline}
    \lvert \Psi(t)  \rangle =  \sum_{\ell \in \mu_\ell} \beta_l e^{-\frac{i}{\hbar}E_{\ell  \bm{E}_0 \lbrace 0 \rbrace} t} \biggl[ e^{-\frac{i}{\hbar} \int_{t_0}^t \Delta_l dt' }\lvert \ell  \bm{E}_0 \lbrace 0 \rbrace \rangle \\ +\sum'_{\ell' \bm{E}_0 \lbrace n \rbrace} \frac{W_{ \ell' \bm{E}_0 \lbrace n \rbrace,\ell  \bm{E}_0 \lbrace 0 \rbrace}}{E_{\ell  \bm{E}_0\lbrace 0 \rbrace} - E_{ \ell' \bm{E}_0 \lbrace n \rbrace}+i\eta} \lvert \ell' \bm{E}_0 \lbrace n \rbrace \rangle  \biggr]
    \label{vacuumexpansion}
\end{multline}

The complex coefficient $\Delta_l$ contains the Lamb shift and the decay rate for the state $\lvert \ell  \bm{E}_0 \lbrace 0 \rbrace \rangle $. Using the wave function in  Eq.~\eqref{vacuumexpansion}, the longitudinal momentum defined in Eq.~\eqref{K} is given by,

\begin{multline}
    \langle \bm{P}_{\mathrm{long}} (t) \rangle = \sum_{\ell,\ell' \in \mu_\ell} \gamma_{\ell \ell'}(t) \sum'_{j  \lbrace n \rbrace}  \langle \ell' \bm{E}_0  \lbrace 0 \rbrace \lvert \frac{e}{c_0} \Delta \bm{A}  \lvert j\bm{E}_0 \lbrace n \rbrace \rangle \\  \frac{W_{j  \bm{E}_0\lbrace n \rbrace, \ell  \bm{E}_0 \lbrace 0 \rbrace}}{E_{\ell \bm{E}_0  \lbrace 0 \rbrace} - E_{j  \bm{E}_0 \lbrace n \rbrace}+i\eta} + c.c
    \label{longitudinal}
\end{multline}
For brevity  we introduced  $\gamma_{\ell \ell'}(t)\equiv \beta_\ell \beta^{*}_{\ell'} \exp({-{i}(E_{\ell \bm{E}_0}-E_{\ell' \bm{E}_0})t/\hbar})$. Because the Stark potential lifts the energy degeneracy
in the subspace $\mu_l$ this coefficient oscillates in time with the Stark frequency.
 Restricting to the creation and annihilation of one virtual photon to stay in the same state we obtain,
\begin{multline}
        \langle \bm{P}_{\mathrm{long}} \rangle = -2\frac{e^2}{c^2_0}  \mbox{Re} \sum_{\ell,\ell' \in \mu_\ell} \gamma_{\ell \ell'}(t)  \sum_{\bm{k}} \mathcal{A}^2_{\bm{k}} \biggl[ \langle \ell' \bm{E}_0 \lvert  (1 - e^{- i \bm{k} \cdot \bm{r}})   \\
      \left( {E_{ \ell \bm{Q}_0 \lbrace 0 \rbrace}-\mathcal{H}_{at}(\bm{p} - \frac{m_2}{M} \hbar \bm{k}, \bm{Q}_0 - \hbar \bm{k}) - \hbar \omega_{\bm{k}} + i\eta}\right)^{-1}  \\
        \Delta_{\bm{k}} \cdot \left(\frac{\bm{p}}{m_1} + \frac{\bm{Q}_0}{M} \right) \lvert \ell, \bm{E}_0 \rangle  + \langle \ell', \bm{E}_0 \lvert  (e^{ i \bm{k} \cdot \bm{r}} - 1) \\
        \left( {E_{ \ell \bm{Q}_0 \lbrace 0 \rbrace}-\mathcal{H}_{at}(\bm{p} + \frac{m_1}{M} \hbar \bm{k}, \bm{Q}_0 - \hbar \bm{k})  - \hbar \omega_{\bm{k}} + i\eta} \right)^{-1} \\
        \Delta_{\bm{k}} \cdot \left(\frac{\bm{p}}{m_2} - \frac{\bm{Q}_0}{M}  \right)  \lvert \ell \bm{E}_0 \rangle \biggr]
        \label{Plongfull}
\end{multline}
In this  lengthy expression, momentum operators have been reintroduced so that the recoil momentum $\hbar \mathbf{k}$ induced by the virtual photon on both proton and electron is highlighted. If $E_{\ell'}<E_{\ell}$ the denominators can actually vanish meaning that real photons can come in provided they have precisely defined energy and momentum.  As $\eta \rightarrow 0$, the delta distribution around this transition gives a finite $\mathbf{k}$-integral and is negligible. Dominant contributions can be identified that simplify  significantly. The terms involving an  exponential $e^{\pm i \bm{k} \cdot \bm{r}}$ do not suffer from a  divergence in the integral over $\mathbf{k}$. Such terms constituted the leading contribution to the Abraham force in the previous section. In this case the leading terms are the matrix elements involving the momentum operators $\pm \bm{p}/m_i + \bm{Q}_0/M$. The $k$-integrals  involving the external momentum $\bm{Q}_0/M$ diverge and can be regularized by mass renormalization \cite{van_tiggelen_qed_2012}.  The divergent parts can be extracted upon taking the large $k$ limit where atomic interactions can be ignored,
\begin{equation}
    \langle \bm{P}_{\mathrm{long}}^{\mathrm{div}} \rangle = \frac{\delta m_1 + \delta m_2}{M} \bm{Q}_0 = \frac{\delta M}{M} \bm{Q}_0
    \label{renormalization}
\end{equation}
with the logarithmically diverging masses
\begin{equation*}
     \delta m_i = \frac{4 \alpha \hbar^2}{3 \pi} \int_0^\infty dk \frac{k}{ \hbar \omega_{\bm{k}} + \frac{\hbar^2 k^2}{2 m_i}}
\end{equation*}
The diverging inertial mass \eqref{renormalization} disappears in the classical momentum $M \dot{\mathbf{R}}$ by redefining the atom mass as $M + \delta M$. The two terms proportional to the internal momenta $\pm \bm{p}/m_i$ have opposite sign and quite remarkably their divergencies at large $k$ cancel.  The dominant contribution comes from virtual, energetic photons for which we can ignore  atomic energies in the denominators. This simplifies to
\begin{eqnarray}
    \langle \bm{P}_{\mathrm{long}} \rangle &=& - \frac{8 \alpha}{3 \pi} \log\left(\frac{m_1}{m_2}\right) \sum_{\ell,\ell' \in \mu_\ell} \gamma_{\ell \ell'}(t) \langle \ell'\bm{E}_0 \lvert \bm{p} \lvert \ell \bm{E}_0 \rangle \nonumber \\
    &=& - \frac{8 \alpha}{3 \pi} \log\left(\frac{m_1}{m_2}\right) \langle \bm{p}(t) \rangle
    \label{Plongleading}
\end{eqnarray}
where we used that
\begin{eqnarray}\label{renorm}
    \frac{\delta m_1}{m_1} - \frac{\delta m_2}{m_2} &=& \frac{4 \alpha \hbar}{3 \pi} \int_0^\infty dk \left[\frac{1}{m_1 c_0 + \frac{\hbar k}{2}} - \frac{1}{m_2 c_0 + \frac{\hbar k}{2}} \right] \nonumber \\
   & =& - \frac{8  \alpha}{3 \pi} \log\left(\frac{m_1}{m_2}\right) <0
\end{eqnarray}
with $m_2 = m_e$ the mass of the electron and $m_1$ the mass of the nucleus. We have written $ \langle \bm{p}(t) \rangle $ for the expectation value of the internal momentum associated with the Stark wave function $ \lvert \Psi(t)  \rangle = \sum_\ell \beta_\ell(t)  \lvert \ell\bm{E}_0  \rangle$. The above integral over photon wave numbers has significant contributions from energetic but still non-relativistic photons with momenta
\begin{equation*}
    \frac{\hbar}{a_n}\approx \frac{\alpha}{n^2} m_{e}c_0  < \hbar k < m_{e} c_0
\end{equation*}
 and for which the present approach should be  valid. Photons with  $k< 1/a_n$ make a positive contribution to  Eq.~(\ref{Plongleading}), but since their wavelength compares to the orbital size $a_n$, atomic energies can  no longer be neglected and small corrections to Eq.~(\ref{Plongleading}) exist. Photons with $k  > m_e c_0^2$ give a finite contribution but should be  given a relativistic approach. This can in fact be done and changes Eq.\eqref{Plongleading} by exactly a factor of 1/4 (Appendix B). We emphasize that this integral must converge in any more advanced description as no classical term exists that can be regularized by mass renormalization.

The next step is to evaluate Eq.~(\ref{Plongleading}) by inserting the Stark wave function in Eq.~\eqref{electricfieldexpansion}. Because $i \hbar \bm{p}/\mu = [ \bm{r},\mathcal{H}_{\mathrm{at}}]$ any contribution to Eq.~\eqref{Plongleading} with $E_{\ell \bm{E}_0} = E_{\ell'\bm{E}_0}$ vanishes.  This is the case for parabolic states that differ only in the sign of the quantum number $m$ and that have the same Stark shift. We can also see that $\langle \ell'' \lvert \bm{p} \rvert \ell \rangle = 0$ for pure parabolic states $\ell$ and $\ell''$ inside  the same subspace $\mu_\ell$. As a result, the second, more complicated part of the Stark wave function \eqref{electricfieldexpansion} conveniently drops out
in the calculation.
We can write $\langle \bm{p}(t) \rangle = (\mu/e) d\langle \bm{d} \rangle /dt $, with $ \bm{d}= e\bm{r} $  the atomic dipole moment. Because for a pure Stark state $\rvert \ell, \bm{E}_0 \rangle$, $\langle \bm{P}_{\mathrm{long}} \rangle$ vanishes, we choose a superposition that corresponds to a finite expectation value of  the angular operator $\mathbf{L}$  in the $xy$-plane. This involves  parabolic states that possess a permanent dipole moment in the $xy$-plane. The angular momentum is not conserved because the electric field exerts a torque on the atom, as described by

\begin{equation}
    \frac{d \langle \bm{L} \rangle }{dt} = \frac{1}{i\hbar} \langle \left[H_{\mathrm{at}}, \bm{L} \right] \rangle = \langle \bm{d} \rangle \times \bm{E}_0
    \label{Torque}
 \end{equation}
This is equivalent to,
 \begin{equation*}
    \frac{d^2 \langle \bm{L} \rangle}{dt^2} \times \bm{E}_0 = -\frac{eE_0^2}{\mu}  \langle \bm{p} \rangle
\end{equation*}
We can check that the angular momentum rotates in the $xy$-plane with the  Stark frequency $\omega_n = \frac{3}{2}n e a_0 E_0/\hbar$, so that $d^2\langle \bm{L} \rangle/dt^2 = - \omega_n^2 \langle \bm{L} \rangle$. We can therefore write Eq.~(\ref{Plongleading}) as,
\begin{equation}\label{Plongfinal}
    \langle \bm{P}_{\mathrm{long}}(t) \rangle =  \frac{3}{\pi} \biggl[  \frac{n^2}{\alpha} \log{\left(\frac{m_1}{m_2}\right)} \biggr] \left( \frac{1}{c_0} \bm{E}_0 \times \langle \bm{m} \rangle (t) \right)
\end{equation}
We inserted $a_0 = \hbar / \alpha m_e c_0$, used  $\bm{m} = -(e/2m_ec_0) \bm{L}$ for the magnetic moment of the Rydberg state and included the relativistic factor of $1/4$.
Expression~(\ref{Plongfinal}) describes an Aharanov-Casher type momentum with a prefactor that is actually much larger than the one put forward in Eq.~(\ref{AB}) although it will turn out to be very small because the allowed external field are constrained by the atomic physics.
 We can now choose a convenient superposition state in the expression~(\eqref{Plongfinal}),

\begin{equation}
\rvert nR  (t)  \rangle= \frac{1}{\sqrt{2}}\biggl[\rvert 0,0,n-1\rangle + e^{-i \omega_n t} \rvert 1,0,n-2\rangle  \biggr]
\label{NR state}
\end{equation}
 For large $n$, this is the superposition of a circular Rydberg state and a nearly circular Rydberg state. This state has the advantage of having a large lifetime needed for the Aharonov-Casher force on the atom to be observable, much larger than the Stark period $1/\omega_n$. If we choose $n=50$ and $E = 1\,  \mbox{V/m}$ we find a typical Stark cycle $1/\omega_n \sim 8\times 10^{-3} \, \mbox{ms}$ whereas the lifetime of the atom is of order $\tau \sim  1\, \mbox{ms}$. For the superposition above the angular momentum rotates along the electric field according to,
\begin{equation*}
\langle \bm{L}(t) \rangle = \hbar \left(\frac{\sqrt{n-1}}{2}\cos(\omega_nt),\frac{\sqrt{n-1}}{2}\sin(\omega_nt),n-\frac{3}{2}\right)
\end{equation*}
so that the momentum oscillates with amplitude
\begin{equation}
|\langle \bm{P}_{\mathrm{long}} \rangle| = \frac{3}{4\pi}n^2\sqrt{n-1} \frac{e a_0 E_0}{c_0}
\end{equation}
frequency $\omega_n$.  This is equivalent to an oscillation
 \begin{equation}\label{dx}
    \Delta x \approx 1/4 \times 4 n \sqrt{n-1}\alpha \frac{m_e}{M}  a_0  \approx \frac{n}{8} \sqrt{n-1} \times 10^{-15}\, \frac{m_p}{M} \, \mathrm{m}
 \end{equation}
of the atom, independent on the applied electric field.  For this choice of parameters, the Aharonov-Casher force on the atom is ${F} \sim 5 \times 10^{-29}\,  \mbox{N}$. This force is still very small but actually a factor $10^3$ larger than the Abraham force observed in Ref.~\citep{rikken_measurement_2011}. In this experiment we estimated $\Delta x = 3 \times 10^{-14}$ m .  The Aharonov-Casher type force found here is directly related to the different mass renormalization by the electromagnetic field quantum vacuum of electron and nucleus.

\section{Conclusions}

In this work we have discussed  QED corrections to the electromagnetic Abraham force  exerted on a highly excited  Rydberg atom, for two different cases. First, we have investigated the Abraham force $\bm{F}= (1/c_0)\langle  \bm{d} \rangle \times \partial_t \bm{B}_0 $ on a singly-excited Rydberg atom subject to a time-dependent, homogeneous magnetic field crossed with respect to a static, homogeneous electric field that induces a dipole moment $\bm{d}$. This extends previous work done on ground state of the Hydrogen atom \citep{van_tiggelen_qed_2012}. The significant degeneration of energy levels and the presence of a large orbital angular momentum are characteristic for the Rydberg states, and are both absent for atomic hydrogen in its ground state. Despite this different picture the dominant contribution stemming from the longitudinal momentum of electromagnetic quantum vacuum (proportional to its vector potential) relative to the classical Abraham force is of same relative order $0.03 \alpha^2$, with $\alpha$ the fine-structure constant, though with opposite sign. This reveals that both classical and quantum force grow as fast as $n^6$ with the principal quantum number $n$ of the Rydberg state. More precisely, the relative QED correction to the Abraham force for $n=50$ is only a factor 4 smaller than the one found for ground state Hydrogen and directed along the direction of the classical force. We conclude that Rydberg atoms behave more or less hydrogen-like with respect to the Abraham force and are therefore good candidates to observe the QED Abraham force, especially because they are more convenient to handle experimentally and have large enough life-times. \\

In the second part of this article we have calculated the QED force on a Rydberg atom in an external static, homogeneous electric field $\bm{E}_0$. If the magnetic moment $\bm{m}$ of the atom is not parallel to the electric field we have found a force
of the Aharonov-Casher type $\bm{F}=(1/c_0)\bm{E}_0 \times d\langle \bm{m} \rangle/dt$.  To our knowledge, this force has no classical equivalent. Our approach reveals a contribution from the quantum vacuum that grows with the principle quantum number $n$. The force is directly related to the mass renormalization in QED that is different for electron and nucleus. Again we conclude that this small force  should be in reach of experimental observation.

\appendix
\section{Transverse momentum of electromagnetic quantum vacuum}
The transverse electromagnetic momentum is associated with real photons with transverse polarization, and given by the operator
\begin{equation}
   \bm{P}_{\mathrm{trans}} = \sum_{\bm{k} \lambda} \hbar \bm{k}  a^{\dagger}_{\bm{k} \lambda} a_{\bm{k} \lambda}
    \label{transversepart}
\end{equation}
We calculate the expectation value for a Stark state $\rvert \ell,\bm{E}_0 \rangle$ and an empty electromagnetic vacuum. We consider the exchange of only one  photon with arbitrary wave number $\mathbf{k}$ with the quantum vacuum as it is slowly turned on and set $\mathbf{Q}_0=0$. We obtain, similar to the approach in section~\ref{section2},
\begin{multline*}
    \langle \bm{P}_{\mathrm{trans}} \rangle = e^2 \exp(2 \eta t /\hbar) \sum_{\ell,\ell' \in \mu_\ell} \gamma_{\ell \ell'}(t) \sum_{\bm{k}} \mathcal{A}^2_{\bm{k}} \, \hbar \bm{k}    \\
    \langle \ell',\bm{E}_0 \lvert \frac{\bm{p}}{m_e}  \cdot \Delta_{\bm{k}}  \frac{1}{\mathcal{H}_{\ell'}+\mathbf{D }-i\eta}\frac{1}{\mathcal{H}_{\ell} +\mathbf{D} +i\eta}\cdot \frac{\bm{p}}{m_e}  \rvert \ell,\bm{E}_0 \rangle
\end{multline*}
where $\mathcal{H}_{\ell} = {\hbar^2 k^2}/{2m_e} + \hbar c_0 k  + \mathcal{H}_{ \mathrm{at} } - E_\ell$ is the operator associated with the change in energy by the emission of the photon and includes photon recoil,  and photon energy and $\mathcal{H}_{ \mathrm{at} }$ is the Hamiltonian introduced in Eq.~({\ref{HAC}). The ``Doppler" operator $\bm{D}=\bm{p}\cdot \hbar \bm{k} /m_e $ discriminates between emitted photons along and opposed to the electron movement without which the $\mathbf{k}$-integral would clearly vanish. Since $\langle p \rangle /m_e \ll c_0$ we can expand both denominators in $\mathbf{D}$ to give
\begin{multline*}
    \langle \bm{P}_{\mathrm{trans}} \rangle = \frac{ 2 \pi e^3 \hbar^3 e^{2 \eta t / \hbar} }{ c_0 m_e^3} \sum_{\ell,\ell' \in \mu_\ell} \gamma_{\ell \ell'}(t) \sum_{\bm{k}} \hat{k}_i k_j (\Delta_{\bm{k}})_{mh} \\
    \langle \ell',\bm{E}_0 \lvert  p_m \biggl[ \frac{1}{\mathcal{H}_{\ell'}-i\eta}p_j\frac{1}{\mathcal{H}_{\ell'}-i\eta}\frac{1}{\mathcal{H}_{\ell}+i\eta} \\
     + \frac{1}{\mathcal{H}_{\ell'}-i\eta}\frac{1}{\mathcal{H}_{\ell}+i\eta}p_j\frac{1}{\mathcal{H}_{\ell}+i\eta}\biggr]
     p_h  \rvert \ell,\bm{E}_0 \rangle
\end{multline*}
If intermediate atom states are inserted as a complete set, contributions appear where the denominators vanish as $\eta \rightarrow 0$. They contribute to spontaneous emission releasing real photons \cite{loudon_quantum_2000} and possibly to $d\langle \bm{P}_{\mathrm{trans}} \rangle /dt \neq 0$ if this  emission is anisotropic for some reason . Here we focuss to $\langle \bm{P}_{\mathrm{trans}} \rangle$ itself corresponding to terms that are finite when $\eta \rightarrow 0$. Large photon energies dominate and we can approximate $\mathcal{H}_{\ell} \approx {\hbar^2 k^2}/{2m_e} + \hbar c_0 k$.
Using $\int_0^\infty dk k^3/\mathcal{H}_{\ell}^3 = m_e/\hbar^4 c_0^2$,
\begin{equation}
    \langle \bm{P}_{\mathrm{trans}} \rangle  \approx \alpha^3 \left(\frac{a_0}{\hbar}\right)^2  \sum_{\ell,\ell' \in \mu_\ell} \gamma_{\ell \ell'}(t) \langle \ell',\bm{E}_0 \lvert  \bm{p}^3 \rvert \ell,\bm{E}_0 \rangle
    \label{transversecontribution}
\end{equation}
If we acknowledge that $\bm{p}^2 \sim \alpha^2 m_e^2 c_0^2$ we see that$\langle \bm{P}_{\mathrm{trans}} \rangle$ is smaller than $\langle \bm{P}_{\mathrm{long}}\rangle$ found in Eq.~(\ref{Plongleading}) by  a factor $\alpha^2$.

section{Relativistic treatment of highly energetic virtual photons}

Our calculations involve the creation and annihilation of virtual photons. Virtual photons with energies $\hbar \omega > m_e c_0^2$ are seen to give a significant contribution. For the highly energetic photons to be correctly treated a relativistic description is needed. Concerning the Abraham force on Rydberg atoms we know that highly energetic virtual photons are negligeable due to the presence of the operator $e^{ \pm i \bm{k}\cdot\bm{r}}$ in the $k$ integrals. It is not the case for the Aharanov-Casher type force on Rydberg atoms. Hence, to get a better estimation of the $k$ integrals in Eq.\eqref{renorm} it is possible to give a relativistic treatment to the problem. The non-relativistic Hamitonian for free particles is replaced by the corresponding relativistic Hamiltonian without spin. We also include the free particle relativistic energy in the interaction Hamiltonian at leading order in the coupling constant. Thus, the atomic hamiltonian and the interaction with quantum vacuum operator are modified as:
\begin{equation}
 H^R_{at} = \sum_{i=1}^2 c_0 \sqrt{p_i^2 + m_i^2 c_0^2} - \frac{e^2}{r}+  \mathcal{H}_S
\end{equation}
\begin{equation}
W^R = \sum_{i=1}^2 \frac{-q_i \bm{p}_i \cdot \bm{A}(\bm{r}_i)}{\sqrt{p_i^2 + m_i^2 c_0^2}}
\end{equation}
where $q_1 = e$ and $q_2 = -e$. The superscript "R" stands for "Relativistic". Expression \eqref{K} for the conserved momentum $\bm{K}$ remains unchanged. We can do the same calculation as we have done with the non-relativistic expressions. The $k$ integral in \eqref{renorm}
\begin{equation}
\int_0^{\infty}dk \biggl[\frac{1}{m_1 c_0 + \frac{\hbar k}{2}} - \frac{1}{m_2 c_0 + \frac{\hbar k}{2}} \biggr] = - \frac{2}{\hbar}\log\left(\frac{m_1}{m_2}\right)
\label{NRkintegral}
\end{equation}
is replaced by
\begin{equation*}
\int_0^{\infty}dk \biggl[\frac{ \hbar k}{\sqrt{\hbar^2 k^2 + m_1^2 c_0^2}(\sqrt{\hbar^2 k^2 + m_1^2 c_0^2}+\hbar k - m_1 c_0)}
\end{equation*}
\begin{equation}
- \frac{ \hbar k}{\sqrt{\hbar^2 k^2 + m_2^2 c_0^2}(\sqrt{\hbar^2 k^2 + m_2^2 c_0^2}+\hbar k - m_2 c_0)} \biggr]
\label{truekintegral}
\end{equation}
The comparison of the integrand of \eqref{NRkintegral} to the one of \eqref{truekintegral} reveals that the error made is mostly concentrated in the momentum window $m_1 c_0 < \hbar k < m_2 c_0$ as expected. The integral \eqref{truekintegral} equals exactly $1/4$ times the integral \eqref{NRkintegral}.

\bibliography{biblio}{}

\end{document}